\documentclass[12pt,draftclsnofoot,onecolumn]{IEEEtran}
\usepackage{url}
\usepackage{ifthen}
\usepackage{cite}
\usepackage{amsmath}
\usepackage{multicol}
\usepackage{graphicx}
\usepackage{amsfonts}
\usepackage{amssymb}
\usepackage{mathrsfs}
\usepackage{tcolorbox}
\usepackage{enumitem}
\usepackage{subfigure}
\usepackage{mdwmath}
\usepackage{mdwtab}
\usepackage{dsfont}
\usepackage[ruled,vlined,linesnumbered]{algorithm2e} 

\DeclareMathOperator*{\argmin}{arg\,min}
\newcommand\smallO{
	\mathchoice
	{{\scriptstyle\mathcal{O}}}
	{{\scriptstyle\mathcal{O}}}
	{{\scriptscriptstyle\mathcal{O}}}
	{\scalebox{.7}{$\scriptscriptstyle\mathcal{O}$}}
}
\newtheorem{remark}{Remark}
\newtheorem{theorem}{Theorem}

\newtheorem{prop}{Proposition}
\newtheorem{defi}{Definition}
\newtheorem{coro}{Corollary}

\def\U{{\boldsymbol{U}}}

\def\u{{\boldsymbol{u}}}

\newenvironment{iarray}{\begin{IEEEeqnarray}{rCl}}{\end{IEEEeqnarray}\ignorespacesafterend}
\newcommand{\dx}{{\sf d}}

\begin{document}

	\title{A Unified Sampling and Scheduling Approach for Status Update in Multiaccess Wireless Networks}
	\author{Zhiyuan Jiang, Sheng Zhou, Zhisheng Niu\\
		Department of Electronic Engineering,\\
		Tsinghua University, Beijing 100084, China\\
		Emails: \{zhiyuan, sheng.zhou. niuzhs\}@tsinghua.edu.cn\\
		\and
		Yu Cheng\\
		Department of Electrical and Computer Engineering,\\
		Illinois Institute of Technology, Chicago 60616, USA\\
		Email: cheng@iit.edu}
	\maketitle

	\begin{abstract}
	Information source sampling and update scheduling have been treated separately in the context of real-time status update for age of information optimization. In this paper, a unified sampling and scheduling ($\mathcal{S}^2$) approach is proposed, focusing on decentralized updates in multiaccess wireless networks. To gain some insights, we first analyze an example consisting of two-state Markov sources, showing that when both optimized, the unified approach outperforms the separate approach significantly in terms of status tracking error by capturing the key status variation. We then generalize to source nodes with random-walk state transitions whose scaling limit is Wiener processes, the closed-form Whittle's index with arbitrary status tracking error functions is obtained and indexability established. Furthermore, a mean-field approach is applied to solve for the decentralized status update design explicitly. In addition to simulation results which validate the optimality of the proposed $\mathcal{S}^2$ scheme and its advantage over the separate approach, a use case of dynamic channel state information (CSI) update is investigated, with CSI generated by a ray-tracing electromagnetic software.
	\end{abstract}
	
	\section{Introduction}
	\label{sec_intro}
	For status update (tracking) in, e.g., latency-sensitive cyber-physical systems, age of information (AoI) \cite{kaul12} is a performance metric that captures the end effect of status observation delay from the destination perspective, and thus enables direct and fair comparisons between systems with differences in, e.g., end-to-end delays, information source sampling rates, throughputs and lossless or lossy designs \cite{roy18}. However, it can be argued that AoI is still an intermediate metric \cite{kam18}, especially in status update systems, whereas the ultimate goal is to optimize the status tracking accuracy at destinations with heterogeneous status information generated remotely by distributed information sources. 
	
	We focus on status update through wireless networks. Existing works in this regard implicitly adopt a \emph{separate} approach---the sampling of information sources and the scheduling of updates/transmissions are treated separately (Fig. \ref{Fig_arch}), with objectives of minimizing the real-time sampling error and AoI respectively \cite{sun17_wiener,sun17,wu13,kam18,kadota18,jiang18_isit,jiang18_itc,kadota16,hsu18,talak18}. Specifically, in terms of information sampling, Ref. \cite{sun17_wiener,sun17,wu13,kam18} consider the scenario that one destination node is remotely tracking the status of one source node, with the objective to minimize, e.g., tracking error \cite{wu13,sun17_wiener} or AoI \cite{sun17}. In this single-source scenario, optimal sampling strategies are obtained under assumptions such as random delay due to channel error \cite{wu13} or limited sampling frequency \cite{sun17_wiener}; these assumptions are made to model the transmissions in wireless networks but are insufficient, due to the considerably more complex behavior of wireless networks. On the other hand, status update scheduling policies usually ignore the structures of information sources, or equivalently, assume the sampling procedure has fully exploited them, and focus on optimizing the AoI \cite{jiang18_isit,jiang18_itc,kadota16,hsu18,kadota18,talak18}. In these works, the information sources are abstracted as status update packets that arrive randomly \cite{jiang18_isit,jiang18_itc,hsu18}, or alternatively as generate-at-will, i.e., active, sources \cite{kadota18,kadota16,talak18}. Scheduling decisions are made to minimize the AoI, irrespective of the status variation---a node may be scheduled even if its status is unchanged since the last schedule.
	
	Despite its simplicity, this separate approach is not necessarily optimal in terms of status tracking accuracy based on a \emph{unified} formulation, and the performance difference is unknown. There have been recent efforts \cite{kam18} towards an effective AoI metric which, compared to AoI, is more directly related to status tracking accuracy in a single-link scenario. The contributions of this paper are summarized as:

	\begin{figure}[!t]
    	\centering
    	\includegraphics[width=0.7\textwidth]{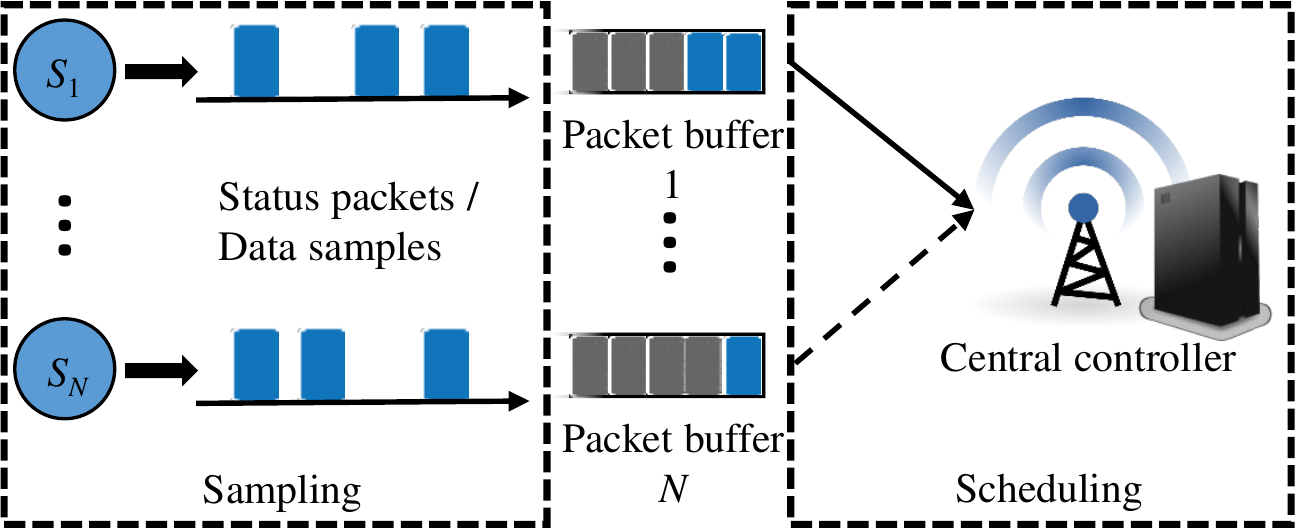}
    	\caption{Status update in a multiaccess wireless network.}
    	\label{Fig_arch}
    \end{figure}

	1) The joint information sampling and transmission scheduling ($\mathcal{S}^2$) problem for multiaccess wireless networks is formulated. Through an example consisting of two-state Markov nodes, it is shown that the optimal $\mathcal{S}^2$ achieves smaller status tracking error than the (near) optimal separate approach. We then consider sources with random-walk state transitions---a discrete version of Wiener processes---and arbitrary error functions; a decentralized near-optimal status update scheme is proposed. Based on extensive simulations, including a use case with channel state information (CSI) sources generated by a ray-tracing software, the proposed scheme shows evident advantage over state-of-the-art separate approaches.
	
	2) From a methodology perspective, to be best of our knowledge, this is the first work that generalizes the Whittle's index to incorporate arbitrary functions which measure the status tracking accuracy, and also the first to apply a mean-filed approach to derive the decentralized policy explicitly. As a special case, the Whittle's index of non-linear AoI (arbitrary reasonable non-linear functions) and the corresponding mean-field based decentralized implementation can be readily given.

	\section{System Model and Preliminaries}
	\label{sec_sm}
	A multiaccess wireless network is considered, where a central controller collects status updates from $N$ distributed source nodes, denoted by $\{S_1,\cdots,S_N\}$. The status updates are conveyed in packets which are transmitted through the wireless multiaccess channel. A time-slotted communication system is considered. In each time slot, only one packet from a source node can be transmitted; a collision happens with more than one simultaneous transmissions. The transmission error probability is $p_{\mathsf{e},n}$ (i.i.d. over nodes and time) if there is no collision. The source nodes only have knowledge of their own statuses in each time slot \cite{sun17_wiener,kam18}.\footnote{In Section \ref{sec_2}, only in order to obtain a performance benchmark, we assume a genie-aided scheduler has knowledge of status changes of all source nodes and thus a centralized scheduling decision is possible.} The packets containing status updates are stored in the buffers at source nodes when waiting to be transmitted. For active sources, the buffers are unnecessary since status packets are generated at will.
	\subsection{General Status Tracking Error Metric for $\mathcal{S}^2$}
	We assume that for source node $S_n$, its status is discrete and belongs to a set $\mathcal{K}_n \triangleq \{\kappa_{n,1},\cdots,\kappa_{n,K_n}\}$. A status $\kappa_{n,k}$ is generic, e.g., sensory data or images, and we assume each status update consumes one packet. We define a general measure of difference between two statuses $\kappa_{n,k}$ and $\kappa_{n,j}$ as $\delta(\kappa_{n,k},\kappa_{n,j})$; examples include $\ell^1$, $\ell^2$ norms when the statuses are described by vectors. Denote the statuses of source node $S_n$ and the destination node under a policy $\pi$ (including information source sampling and transmission scheduling) at time $t$ as $s_n(t)$ and $\hat{s}_{n,\pi}(t)$ respectively, and $s_n(t)$, $\hat{s}_{n,\pi}(t) \in \mathcal{K}_n$. The $T$-horizon time-average remote status tracking error can be expressed as $\overline{\Delta}_{n,\pi}^{(T)} \triangleq \frac{1}{T}\sum_{t=0}^{T-1} \delta(s_n(t),\hat{s}_{n,\pi}(t))$. In particular, we are interested in the infinite-horizon regime where $T \to \infty$. Denote the weighted long-time-average remote status tracking error of all sources as $\overline{\Delta}_{\pi} \triangleq \limsup_{T \to \infty} \frac{1}{N} \sum_{n=1}^{N} w_n \overline{\Delta}_{n,\pi}^{(T)}$, where $[w_1,\cdots,w_N]$ is a weight vector denoting the importance of each source node. The objective is therefore to seek for a policy $\pi$ that minimizes the time-average error, i.e.,
	\begin{equation}
	\label{obj}
	    \min_{\pi} \overline{\Delta}_{\pi},\textrm{ subject to }\sum_{n=1}^N u_n(t)\le1,\, \forall t,
	\end{equation}
	where $u_n(t)=1$ denotes the source node $S_n$ transmits successfully during time slot $t$; otherwise $u_n(t)=0$.

	\subsection{Conventional Separate Approach}
	In a separate sampling and scheduling approach, the information sampling design concerns with the optimal sampling time of a single source to minimize the remote tracking error, with some assumptions on the communication part, e.g., fixed delay \cite{kam18}, random delay \cite{sun17_wiener} or unreliable channel \cite{wu13}; additional constraints include limited sampling frequency \cite{sun17_wiener}. Afterwards, the transmission scheduling decision is based on optimizing the AoI \cite{kadota18,hsu18,jiang18_itc,jiang18_isit} which is defined as follows. The $T$-horizon time-average AoI of source node $S_n$ is $\overline{h}_{n,\pi}^{(T)} \triangleq \frac{1}{T}\sum_{t=0}^{T-1} h_{n,\pi}(t)$, where $h_{n,\pi}(t)$ denotes the AoI reported by $S_n$ at the $t$-th time slot under policy $\pi$ and $h_{n,\pi}(t) \triangleq t - \mu_{n,\pi}(t)$, where $\mu_{n,\pi}(t)$ denotes the sampling time of the newest status received at the destination until time $t$. The weighted long-time-average AoI is defined as $\overline{h}_{\pi} \triangleq \limsup_{T \to \infty} \frac{1}{N} \sum_{n=1}^N w_n \overline{h}_{n,\pi}^{(T)}$. 
	
	The rationality of this separate approach is that by optimizing the AoI, the tracking error can be minimized consequently by assuming the tracking error grows with the staleness of the information obtained. However, this assumption needs to be reconsidered with generic status variation. The motivating example introduced in the following section will show that this separate approach is not optimal and can be improved by a unified sampling and scheduling approach. 
    
	\section{Status Tracking Error Analysis for a Network of Two-State Markov Sources}
	\label{sec_2}
	Consider a network with $N$ independent Markov information sources, each with two symmetrical states $\{0,1\}$ and state change probability of $p_n \triangleq p_{n,0 \rightarrow 1}=p_{n,1\rightarrow 0}$.\footnote{We only consider $p_{n} \le 0.5$ since a symmetrical two-state Markov source with $p_n > 0.5$ can be regarded as the modulo-$2$ superposition of a constant source with alternative $0$ and $1$ output and a source with change rate $1-p_n$.} We assume reliable channels in this section and later on generalize in Section \ref{sec_whittle}.
	The performance metric is the long-time sample average error probability which converges to the stochastic average denoted by $\Pr(\varepsilon_n|\pi)$ given policy $\pi$, i.e., $\delta(s_n(t),\hat{s}_{n,\pi}(t)) = \mathds{1}_{\{s_n(t)\neq\hat{s}_{n,\pi}(t)\}}$, $\Pr(\varepsilon_n|\pi) = \limsup_{T \to \infty} \frac{1}{T}\sum_{t=0}^{T-1} \mathds{1}_{\{s_n(t)\neq\hat{s}_{n,\pi}(t)\}}$. We further assume in this section that $w_i=1$, $\forall i \in \{1,\cdots,N\}$, namely all nodes have equal weights. The objective function in \eqref{obj} is therefore $\overline{\Delta}_{\pi} = \frac{1}{N} \sum_{n=1}^{N} \Pr(\varepsilon_n|\pi)$.
	
	\subsection{Optimal $\mathcal{S}^2$ for Two-State Sources}
	In this section, we assume that there is a genie coordinating the status update decisions with all the information to obtain the optimal performance as a benchmark; this assumption is unrealistic and will be removed in the next section considering decentralized status update with only local information. The Markov decision process (MDP) is formulated to obtain the optimum, based on which the system state is $\mathcal{B}(t) \triangleq \{(s_1(t),\hat{s}_1(t)),\cdots,(s_N(t),\hat{s}_N(t))\}$ where $s_n,\hat{s}_n \in \{0,1\}$. The transition probability with action $u_n(t)=\mathds{1}_{\{n=m\}}$, $\forall n \in \{1,\cdots,N\}$ is 
	\begin{equation}
	\Pr\left\{\mathcal{B}(t) \to \mathcal{B}(t+1)\right\} = \prod_{n \in \Psi_1} p_n \prod_{i \in \Psi_2} (1-p_i), \nonumber
	\end{equation}
	where $\Psi_1$ is the set of source nodes whose statuses change at time slot $t$, and $\Psi_2$ contains those unchanged. The statuses at time $t+1$ at the central controller remain the same as time $t$, i.e., $\hat{s}_n(t+1) = \hat{s}_n(t)$, $\forall n \neq m$, except for node $m$ who is updated to $\hat{s}_m(t+1) = {s}_m(t)$. Define the $T$-horizon total cost function as (a constant factor of $\frac{1}{N}$ is omitted)
	\begin{iarray}
		\label{minE}
        \min_{u_n(t),\,n \in \{1,\cdots,N\}} \mathbb{E}\left[\sum_{t=0}^{T-1} \sum_{n=1}^N \mathds{1}_{\{s_n(t)\neq\hat{s}_{n,\pi}(t)\}} \right].
	\end{iarray}	
	The optimal policy is as follows.
	\begin{tcolorbox}
	\begin{theorem}[Optimal $\mathcal{S}^2$ for Two-State Markov Sources]
	\label{thm1}
	The optimal policy for the MDP is at time $t$, node $S_{n_\mathsf{opt}}$ updates its current status where
	\begin{iarray}
	\label{nc}
	n_\mathsf{opt} &=& \argmin_{n\in \chi(t)} p_n,
	\end{iarray}
	and $\chi(t)$ is the set of nodes with $s_n(t)\neq\hat{s}_{n}(t)$, and ties are broken arbitrarily. 
	\end{theorem}
	\end{tcolorbox}
	\begin{IEEEproof}
	The optimal policy is to schedule the source node that has error, and meanwhile is the least likely to change (avoiding repetitive and hence wasted update). The detailed proof is given in Appendix \ref{app_thm1}.
	\end{IEEEproof}
	\begin{remark}
	We have derived the optimal policy for the two-state system with a finite horizon, which is considered to be the heavy-lifting of the proof. The generalization to infinite horizon follows standard methods, cf. \cite[Theorem 2]{lu84}. Although significantly simplified, the considered scenario presents itself when considering a network of nodes for e.g., anomaly detection which has two states: normal and alert. $\hfill\square$
	\end{remark}
	\subsection{Near-Optimal Separate Approach}
	In this subsection, we aim to obtain the optimal separate policy and compare its performance with that of Theorem \ref{thm1}, in order to show whether $\mathcal{S}^2$ is beneficial in this scenario. 
	
	To begin with, the sampling strategy which is independent with the transmission scheduling is considered. The \emph{sample-at-change} sampler, i.e., sampling the source whenever there is a state change, is clearly the optimal lossless sampler in terms of minimum sampling times. Moreover, since there are only two states, the optimal sampling strategy in \cite{sun17_wiener}, which is based on a threshold of the status change and proved optimal considering Wiener sources, reduces to sample-at-change. In general, we remark that the optimal sampling strategy is related to assumptions on communications, e.g., random delay and erasure channel. However, it is very difficult to model the overlay wireless network perfectly. Hence, we adopt the sample-at-change strategy which is also used in \cite{kam18}.
	
	According to the separate approach, the sampling strategy deals with the information source structure and the scheduling policy is designed to minimize the AoI. Based on the sample-at-change sampler, the sampled packets arrive at the buffer of $S_n$ based on a Bernoulli distribution with parameter $p_n$. There are rich existing works on the scheduling policy for AoI optimization in this system setting \cite{kadota18,hsu18,jiang18_isit,jiang18_itc}. It has been shown that although the optimal scheduling policy seems elusive, a closed-form Whittle's index based policy is obtained in \cite[Theorem 3]{jiang18_itc}. Thereby, each node only retains the most up-to-date packet, and the node with the highest index is scheduled. The index is given by $m(a_n,b_n,p_n)=\frac{1}{2} x_n^2 + \left(\frac{1}{p_n}-\frac{1}{2}\right) x_n$, if $b_n > \frac{p_n}{2}(a_n^2-a_n)+a_n$; and $\frac{b_n}{p_n}$ otherwise, where $x_n\triangleq \frac{b_n + \frac{a_n(a_n-1)}{2}p_n}{1-p_n+a_n p_n}$, the age of the packet at buffer of $S_n$ is denoted by $a_n$, the current AoI of $S_n$ is denoted by $h_n$ and $b_n \triangleq h_n - a_n$. It is well-known that the Whittle's index based policy can achieve near-optimal performance of restless multi-armed bandit problems, especially with a large number of source nodes \cite{web90}. Therefore, we claim that the index policy is close to optimal to minimize the AoI of the system. 
    
    Now we have obtained the optimal $\mathcal{S}^2$ policy and a near-optimal separate policy in a network with independent $0$-$1$ Markov sources, the status tracking error performance comparisons are in order. 
    \subsection{Comparisons between $\mathcal{S}^2$ and Separate Approach}
    \begin{figure}[!t]
    	\centering
    	\includegraphics[width=0.7\textwidth]{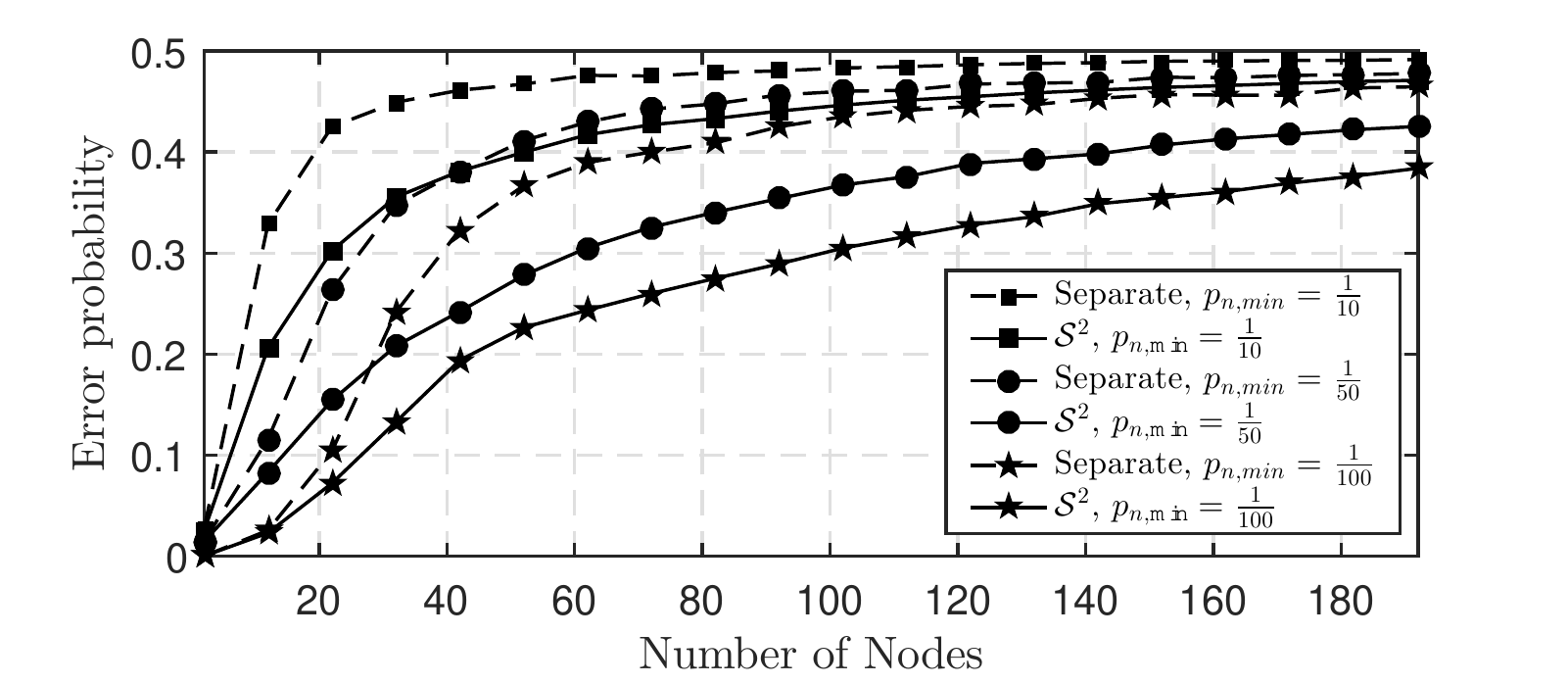}
    	\caption{Tracking error comparisons between optimal $\mathcal{S}^2$ and near-optimal separate approach based on the two-state Markov sources setting.}
    	\label{Fig_error2}
    \end{figure}
    In Fig. \ref{Fig_error2}, we compare the status tracking error performances of the two-state Markov source network with optimal $\mathcal{S}^2$ derived in Theorem \ref{thm1}, and the near-optimal Whittle's index policy based separate sampling and scheduling approach described in the last subsection. The transition probability $p_n$ is uniformly distributed in $[p_{n,\mathsf{min}},\frac{1}{2}]$. The simulation runs for $10^6$ time slots. It is observed that the performance gap between $\mathcal{S}^2$ and the separate approach is evident. The reason is that the separate approach, although optimizing the status update timeliness, does not account for the real status change, i.e., instead of being event-triggered, the separate approach is timing-triggered. In this specific case, the separate approach may squander status update opportunities on nodes with large AoI but correct statuses due to the back-and-forth state transition of the two-state Markov chain. In contrast, the optimal $\mathcal{S}^2$ selects the node with both incorrect status and smallest $p_n$ (least likely to change back). 
    
    Having witnessed the potential benefit by $\mathcal{S}^2$, let us turn to a more general scenario and consider more practical issues.
    
    \section{Decentralized Event-Triggered Status Update for Nodes with Random-Walk State Transitions}
    \label{sec_randomwalk}
    One can immediately realize that the optimal $\mathcal{S}^2$ proposed in Theorem \ref{thm1} is not feasible, expect for illustration purposes, in the sense that it requires a genie to inform the scheduler about the set of nodes where events (status changes) happen at the considered time. Moreover, even the assumption of a central scheduler to make scheduling decisions is questionable, considering the prohibitive signalling overhead in the massive Internet-of-Things (IoT) system. In summary, there are two distinct challenges in designing event-triggered status update in a multiaccess wireless network:
    
    1) How to measure the \emph{event importance} for status update such that by scheduling nodes with important updates, the overall status tracking accuracy is improved?
    
    2) How to design a \emph{decentralized} status update scheme based on the event importance?

    A natural idea to measure the event importance is by measuring the status change amplitude; however, problems arise in the case with heterogeneous nodes, e.g., different error functions due to different nodes' sensitivities towards error. To address this issue, we adopt a Whittle's index-based solution. 
    
    A high-level overview of the proposed scheme, i.e., event-triggered status update (ETSU), is as follows. The Whittle's index, which is denoted by $I_n(t)$ for node $n$ at time $t$, is calculated by each source node based solely upon its local information, e.g., status change; it is therefore adopted to measure the status change importance and triggers the status update. Each node maps its index to a transmission probability $p_n(t)$ based on a pre-defined function $\operatorname{\Psi}(I_n(t))$ which is public and identical among nodes (to be specified later), i.e.,
    \begin{equation}
    \label{map}
    p_n(t) = \operatorname{\Psi}(I_n(t)), \, \forall n\in \{1,\cdots,N\}.
    \end{equation}
    The nodes undergo a contention period based on their distinct transmission probabilities \eqref{map} and the winner transmits a status update packet. Intuitively, a higher index, indicating a higher importance, should be mapped to a higher transmission probability. The methodology is described here.
    \begin{tcolorbox}[title=ETSU]
    \textbf{Contention period}: \\
	\noindent Node $S_n$ transmits with probability $p_n(t) = \operatorname{\Psi}(I_n(t))$.\\
	\textbf{Transmission or collision time slot}:\\
	\noindent In case of collision, the central controller feeds back a NACK; otherwise ACK.\\
	Go to the contention period.		
    \end{tcolorbox}
    
    ETSU can be viewed as a prioritized $p$-persistent carrier-sense multiple access (CSMA) scheme, and has been proposed in \cite{jiang18_itc} to address AoI optimization in wireless uplinks. However, the challenges here are to design $I_n(t)$ to accommodate event importance instead of AoI in \cite{jiang18_itc}, and to design the mapping $\operatorname{\Psi}(\cdot)$ explicitly for which we adopt a novel mean-field approach. In what follows, we tackle these challenges one-by-one. 
    \subsection{Whittle's Index for Nodes with Random-Walk Transitions}
    \label{sec_whittle}
    Two assumptions on state transitions and status tracking error metric are in order before diving into the index derivation.

    1) We consider source nodes with state transitions modeled as random walk on a one-dimensional line. Formally, $s_n(t+1)=s_n(t)+a_n(t)$, where
    \begin{equation}
        \Pr\{a_n(t)=i|s_n(t)\}=\Pr\{a_n(t)=i\}=q_{n,i},
    \end{equation} 
    and $q_{n,1}=0.5$, $q_{n,-1}=0.5$, $\forall n$. Hence the status evolution of each node is a random walk with i.i.d. (among nodes) and time-homogeneous increments.
    
    2) The status tracking error functions $\delta_n(s_{n}(t),\hat{s}_{n}(t))$, $\forall n \in \{1,\cdots,N\}$, is only related the absolute status difference, i.e., $\delta_n(s_{n}(t),\hat{s}_{n}(t)) = \delta_n\left(|s_{n}(t)-\hat{s}_{n}(t)|\right)$, and satisfy: $\forall n\in \{1,\cdots,N\}$, 
        \begin{iarray}  
        \label{de}
        \delta_n(0)=0 \le \delta_n(1) \le\cdots\le \delta_n(d)\le\cdots,
        \end{iarray}
    and there exists $D>0$, $\forall d \ge D$, $\delta_n(d)>0$.

    The random-walk state transitions capture the essence of many real-world physical phenomenons, e.g., the stock price and the path traced by a foraging animal; the Wiener process considered in the literature \cite{sun17_wiener} is the scaling limit of a random walk with very small steps. Note that the state values $s_{n}$ should be viewed as normalized status values, and the end effect of heterogeneous status values, e.g., temperature status ranging in $[-30,30]$ and an image pixel in $[0,255]$, is captured by distinct error functions $\delta_n(\cdot)$. Several extensions to the random walk assumption are considered trivial and thus not treated specifically: Random walk with different step sizes which, as mentioned before, can be treated by formulating different error functions $\delta_n(\cdot)$; Asymmetrical random walk where, say $q_{n,1}>0.5$, can be treated with adding a constant drift of $(2q_{n,1}-1)\tau$, where $\tau$ denotes the last update time, to the tracking status $\hat{s}_n(t)$; For random walk with stay probability, i.e., $q_{n,0}\neq0$ which can be different among nodes, the corresponding index is the one without stay probability multiplied by a factor of $1/(1-q_{n,0})$ by the standard Markov chain uniformization technique \cite[Chapter 6.4]{gal12}. One other reason to consider random walks is that state transitions of information sources in real world are difficult to model and thus often unknown, whereas random walks offer an approximation.
    
    The second assumption, together with the random walk assumption, essentially makes the problem \emph{state-homogeneous}, i.e., the status tracking difference is sufficient statistics. Therefore, the system state can be simplified to be one-dimensional, i.e., $(s_n(t),\hat{s}_n(t)) \Leftrightarrow (d_n(t)\triangleq|s_n(t)-\hat{s}_n(t)|)$. The state-homogeneous error functions are still considered quite general since a wide range of metrics, e.g., $\ell^1$, $\ell^2$ norms and threshold-type error functions, can be included. The requirement of non-decreasing of $\delta_n(d)$ is also reasonable, considering a larger status difference should have a larger status tracking error.
    
    The Whittle's index is a technique to deal with the restless multi-armed bandit (RMAB) problem which usually suffers from the curse of dimensionality when considering a large number of arms. Our status update problem clearly fits into the RMAB framework since the status update decisions for nodes are binary, like pulling arms of bandit machines, and the status changes even if the node is not scheduled i.e., restless. The philosophy of the Whittle's index is to decompose the $N$-dimensional RMAB into $N$ one-dimensional MDP problems; solving each one-dimensional problem is equivalent to comparing each bandit machine to a machine with constant payoff (or cost) $m_n$, which is also equivalent to solving a relaxed Lagrange dual problem \cite{web90}. By this methodology, the index $m_n$ can be viewed as a measure of the event importance optimized for multi-node status update. The challenge when applying the Whittle's index is that it is only defined for a subset of RMAB problems which are indexable.
    
    \begin{defi}[Indexability]
    For the decomposed problem, given auxiliary costs $m_1$ and $m_2$, and the sets of states that the optimal action is to idle denoted by $\Pi_{m_1}$ and $\Pi_{m_2}$ respectively, the RMAB is indexable if $\forall 0 \le m_1 < m_2,\, \Pi_{m_1} \subseteq  \Pi_{m_2}$, and $\Pi_{0} =  \emptyset$; for $\Pi_{+\infty}$ is the entire state space. $\hfill\square$
    \end{defi}
    
    Several existing works \cite{kadota16,kadota18,hsu18,jiang18_itc,singh15} have adopted the Whittle's index. Hence we omit the index formulation and problem decomposition, diving directly into solving the decomposed problems which are described as follows (the node and time indices are omitted for brevity). 
    \begin{equation}
    \label{c2go}
    f(d) + \hat{J}^*  =  \min \left\{ \begin{array}{l}
    \delta(d) + \frac{f(d+1)}{2} + \frac{f(d-1)}{2} ,\\
    m + \frac{f(1)}{2} + \frac{f(0)}{2}
    \end{array} \right\}, 
    \end{equation}
    with $d \ge 1$. Considering the long-time average cost MDP formulation, the relative cost function is denoted by $f(d)$, the optimal average cost is denoted by $\hat{J}^*$, and the auxiliary constant cost in the Whittle's index methodology is denoted by $m$. We can prescribe $f(0)=0$ and the state-$0$ will transit to state-$1$ with probability $0.5$; otherwise stays at state-$0$.  
    \begin{tcolorbox}
    \begin{theorem}[Whittle's Index for Random-Walk Information Sources]
    \label{thm2}
    The Whittle's index, denoted by $I_{\mathsf{RW},n}(d_n)$ with status difference $d_n=|s_n-\hat{s}_n|$, is
    \begin{iarray}
    \label{whittle}
    I_{\mathsf{RW},n}(d_n) = w_n \sum_{i=1}^{d_n}\left(2i-d_n\right)\delta_n(i).
    \end{iarray}
    The problem is indexable. 
    \end{theorem}
    \end{tcolorbox}
    \begin{IEEEproof}
    A sketch of the proof is given in Appendix \ref{app_thm2}.
    \end{IEEEproof}
    \begin{coro}[Index for Unreliable Channels]
    \label{coro1}
    Assuming the transmission error probability is $p_{\mathsf{e},n}$ (no collision), the corresponding Whittle's index, denoted by $\hat{I}_{\mathsf{RW},n}(d_n,p_{\mathsf{e},n})$, is the index without transmission error multiplied by $1-p_{\mathsf{e},n}$, i.e., $\hat {I}_{\mathsf{RW},n}(d_n,p_{\mathsf{e},n}) \overset{p_{\mathsf{e},n} \to 0}{\longrightarrow} I_{\mathsf{RW},n}(d_n) (1-p_{\mathsf{e},n}) + \smallO(1)$. $\hfill\square$
    \end{coro}
    \begin{IEEEproof}
    This result has an intuitive explanation: For a node with transmission error probability $p_{\mathsf{e},n}$, it takes approximately $1/(1-p_{\mathsf{e},n})$ times to reach a successful transmission whose equivalent service charge (system is willing to pay) is $I_{\mathsf{RW},n}(d_n)$, and hence the corresponding index with error is the one without transmission error divided by $1/(1-p_{\mathsf{e},n})$ times of trials. However, the exact expression with general value of $p_{\mathsf{e},n}$ is more complicated and cumbersome without insights. Therefore, we give a sketch of the proof in Appendix \ref{app_coro1} for the case with $p_{\mathsf{e},n} \to 0$ which has this clear structure.
    \end{IEEEproof}
    
    The Whittle's index policy which, at each time, updates the node with the largest index has been shown to have a strong performance for a wide range of RMAB problems \cite{jiang18_itc,kadota18,hsu18,singh15}, especially in scenarios with a large number of nodes. The results in Theorem \ref{thm2} and Corollary \ref{coro1} are, as far as we know, the first results that establish closed-form Whittle's indices and their indexability for general and heterogeneous cost functions $\delta_n(\cdot)$.
    \subsection{Optimality of Whittle's Index with Homogeneous Nodes}
    The Whittle's index policy is in fact optimal for a specific scenario where nodes are homogeneous, i.e., identical error function and identical transmission error. Similar observations that the index policy is optimal in symmetrical networks are found in, e.g., \cite{kadota18,singh15}, however in different contexts where only identical transmission error is considered. 
    \begin{prop}
    \label{thm3}
    Assuming random-walk state transitions and $\delta_n(\cdot)=\delta(\cdot)$, $w_n=1$, $p_{n,\mathsf{e}}=p_{\mathsf{e}}$, $\forall n \in \{1,\cdots,N\}$, the index policy based on \eqref{whittle} is optimal. $\hfill\square$
    \end{prop}
    \begin{IEEEproof}
    The proof technique is similar with the proof of Theorem \ref{thm1}. Basically, the index policy updates the node with the largest status difference in this case, which is also the myopic policy based on the MDP formulation. Again, based on backwards induction, it can be shown that the myopic policy is optimal. The details are omitted for brevity.
    \end{IEEEproof}
    
    \subsection{Design of $\Psi(\cdot)$: A Mean-Field Approach}
    The design of the mapping from index, representing the event importance, to transmission probability in a system with a large number of nodes can be formulated into a mean-field control problem \cite{lasry07}. The key idea is that when $N \to \infty$, under some conditions specified later, the control of all the nodes can be simplified to the control of the mean-field.
    \begin{defi}[Mean-Field]
    The mean field is defined as the probability distribution of the states over the set of nodes, i.e., $\mathsf{F}(d,t) \triangleq \lim_{N \to \infty}\frac{1}{N} \sum_{n=1}^N \mathds{1}_{\{d_n(t)=d\}}$. $\hfill\square$
    \end{defi}
    
    The general assumptions that guarantee the convergence of the optimal control to the mean-field control \cite{lasry07} include interchangeability of nodes, and interaction with the mean-field for each node. For ease of exposition, we consider one class of homogeneous nodes (identical error functions) in this subsection, such that the former assumption is satisfied; extensions to multiple classes of nodes are straightforward by treating nodes in each class as a mean-field. The latter assumption is also satisfied since every node only concerns with the statistical distribution of other nodes, instead of the state of each individual node. 
    
    Unlike the conventional mean-field approach that adopts the Hamilton-Jacobi-Bellman (HJB) equation \cite{lasry07} to account for the system dynamics, the Whittle's index solution derived in Theorem \ref{thm3} has significantly simplify the control, namely the objective at each time slot is only to schedule the node with the largest index (in the homogeneous case, it is equivalent to the largest $d$) by considering the mean-field of the states. A schematic of the approach is specified as
    \begin{equation}
        D_{\mathsf{th}} \mapsto F_n(D_{\mathsf{th}}) \overset{(a)}{\mapsto} \overline{F}(D_{\mathsf{th}}) \overset{\overline{F}(D_{\mathsf{th}}^*)=1-\nu}{\mapsto} D_{\mathsf{th}}^*,\nonumber
    \end{equation}
    which can be explained as follows. First select an index threshold $I_{\mathsf{th}}(D_{\mathsf{th}})$, and any node with index above the threshold transmits with probability $p_\mathsf{tx}$; this produces a stationary distribution of each node $ F_n(D_{\mathsf{th}})$. The equivalence of $(a)$ stems from the mean-field assumption, i.e., the state distribution over the set of nodes approaches the stationary distribution of one node in the large system limit. Finally, the index threshold is derived which allows nodes with indices larger than the top $\nu$-fraction of all nodes to transmit with the same probability $p_\mathsf{tx}$. Note that ideally, the maximum index of the mean-field is scheduled; however such an approach is unstable given a continuous mean-field distribution. Therefore we adopt a mapping function that lets the top $\nu$-fraction of nodes compete:
    \begin{equation}
    \label{mapping}
        \Psi(I) = \left\{\,
            \begin{IEEEeqnarraybox}[][c]{l?l}
            \IEEEstrut	
            p_\mathsf{tx}, &\textrm{if } I \ge I_{\mathsf{th}}; \\
        	0,&\textrm{otherwise}.
        	\IEEEstrut
        	\end{IEEEeqnarraybox}
        	\right. 
    \end{equation}
    The following theorem gives closed-form expressions of $I_{\mathsf{th}}$ and $p_\mathsf{tx}$ for a general Markov chains where the state transition given $D_{\mathsf{th}}$ can be described in the below figure.
    \begin{figure}[!h]
    	\centering
    	\includegraphics[width=0.8\textwidth]{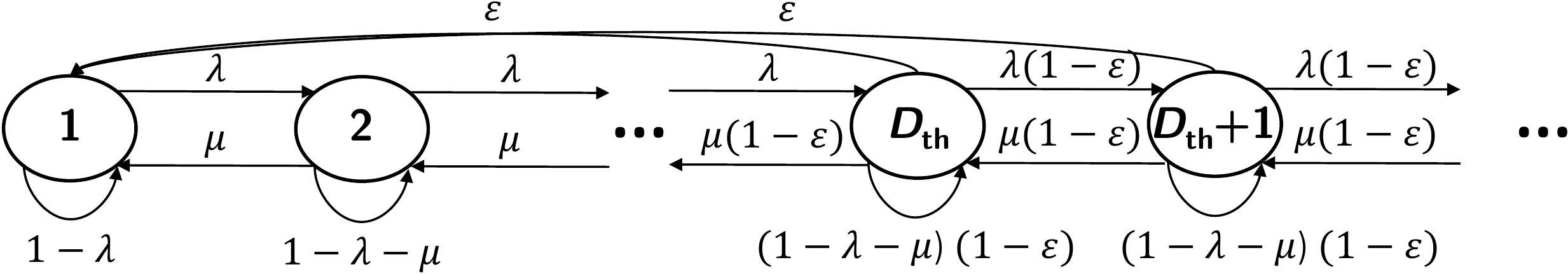}
    	\label{Fig_mf}
    \end{figure}
    The random walk considered in Section \ref{sec_whittle} is a special case with $\lambda=\mu=0.5$. Note that the transition in only among adjacent states; this is not a restriction when sufficiently small time slots are considered. The probability of successful transmission when $D\ge D_{\mathsf{th}}$ is $\varepsilon=\frac{1}{\nu N}$, i.e., the top $\nu$-fraction of nodes ($\nu N$ nodes) get equal transmission probability. 
    \begin{tcolorbox}
    \begin{theorem}[Optimal Mean-Field Control]
    \label{thm_mf}
    When $N\to \infty$, the index threshold $I_{\mathsf{th}}$ that allows the top $\nu$-fraction of nodes to transmit is given by the index expression, e.g., \eqref{whittle}, with $d$ being the unique solution of the equation
    \begin{equation}
    \label{48}
        \nu^{-1} = 
        \left\{ \begin{array}{ll} \frac{\left(\frac{\lambda}{\mu}\right)^{1-d}-\frac{\lambda}{\mu}}{1-\frac{\lambda}{\mu}}\left(\frac{1}{\lambda\beta}-\frac{\varepsilon}{\lambda-\mu}\right)+\frac{\varepsilon d}{\lambda-\mu}+1,&\lambda\neq \mu, \\
        \frac{d}{\lambda\beta}+\frac{\varepsilon(d-1)d}{2\lambda}+1,&\lambda=\mu,
        \end{array} \right.
    \end{equation}
    where $\varepsilon=\frac{1}{\nu N}$, $\beta = \boldsymbol{e}_1^{\mathsf{T}}\left(\boldsymbol{I}-\hat{\boldsymbol{P}}\right)^{-1} \boldsymbol{1}$ (see \eqref{12}). The transmission probability $p_{\mathsf{tx}}$ of the top $\nu$-fraction of nodes is given by the unique solution of the equation
    \begin{equation}
    \label{41}
        (1-p_{\mathsf{tx}})^{\nu N} = \frac{t_{\mathsf{slot}}}{t_\mathsf{c}}\left(\nu N p_{\mathsf{tx}} + (1-p_{\mathsf{tx}})^{\nu N} -1 \right),
    \end{equation}
    where $t_\mathsf{slot}$ is the duration of one time slot and $t_\mathsf{c}$ is the duration of one contention slot.
    \end{theorem}
    \end{tcolorbox}
    \begin{IEEEproof}
    See Appendix \ref{app_thm_mf}.
    \end{IEEEproof}
    \begin{coro}[Closed-Form Solution for Random-Walk Nodes]
    \label{coro_mf}
    When $\lambda=\mu$, the solution of \eqref{48} is
    \begin{equation}
        d = \frac{1}{2}-\frac{\nu N}{\beta}+2\sqrt{\frac{\nu^2 N^2}{\beta^2} - \frac{\nu N}{\beta} + \frac{1}{4} - 2 \lambda \nu N (1-\nu^{-1})}, \nonumber
    \end{equation}
    \begin{equation}
        \beta = \frac{1+{a/2b}+\sqrt{{a^2/4b^2}-1}}{a+2b},\nonumber
    \end{equation}
    where $a \triangleq 2\lambda(1-\frac{1}{\nu N})+\frac{1}{\nu N}$, and $b \triangleq -\lambda\left(1-\frac{1}{\nu N}\right)$. $\hfill\square$
    \begin{IEEEproof}
    The proof is based on an inversion lemma of Toeplitz tridiagonal matrices given in \cite[Theorem 3.3]{kamp89}. The details are omitted due to lack of space.
    \end{IEEEproof}
    \end{coro}
    
    The above methodology can be readily applied to derive the decentralized mean-field control for non-linear AoI cost (arbitrary function of AoI satisfying the non-decreasing assumption) minimization problem \cite{kosta17}. Define the AoI cost function as $g(h)$ and consider the generate-at-will sources whereby each update drops the AoI to one. The following corollary gives the mean-field control based on Whittle's index and scheduling design of ETSU.
    \begin{tcolorbox}
    \begin{coro}[Mean-Field Control for Non-Linear AoI]
    \label{coro_mf_aoi}
    The index threshold with arbitrary AoI cost function $g(h)$, $\forall h
    \ge 1$ is given by
    \begin{equation}
        I_{\mathsf{th}}=\sum_{h=1}^{H_{\mathsf{th}}}\left(g\left(H_{\mathsf{th}}\right)-g(h)\right)(1-p_{\mathsf{e}}), \nonumber
    \end{equation}
    where $H_{\mathsf{th}} = \frac{(1-\nu)N}{1-p_{\mathsf{e}}}$, the transmission error probability is $p_{\mathsf{e}}$, and by setting $\nu$ we let the top $\nu$-fraction of nodes to transmit with probability given by \eqref{41}.
    \end{coro}
    \end{tcolorbox}
    We have thus far addressed the two issues proposed at the beginning of the section, by applying Theorem \ref{thm2} to decide the event importance and Theorem \ref{thm_mf} based on mean-field approach to determine the importance-to-transmission-probability mapping of $\Psi(\cdot)$ in \eqref{mapping}.
    \subsection{Performance Evaluations}
\begin{figure}[!t]
        \centering    
        \label{fig_index}
        \subfigure[]{\includegraphics[width=0.46\textwidth]{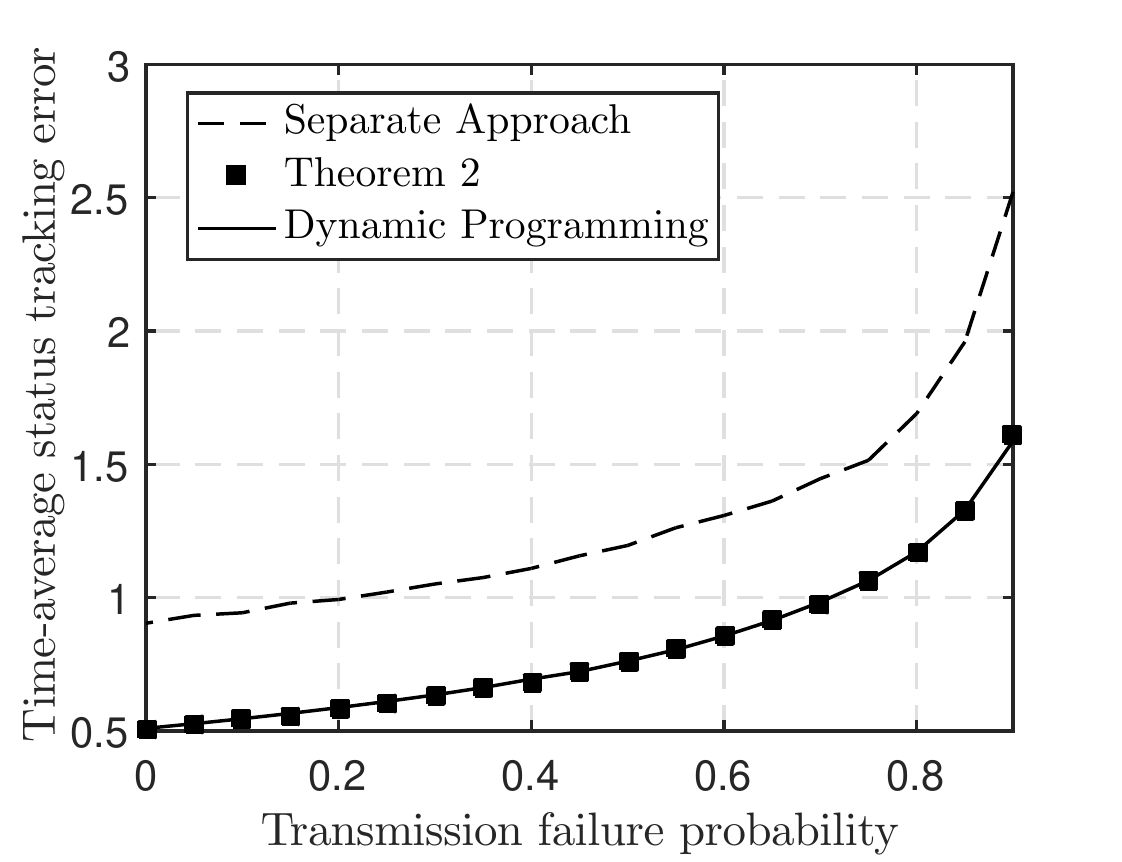}}
        \label{fig_ipra}
        \subfigure[]{\includegraphics[width=0.46\textwidth]{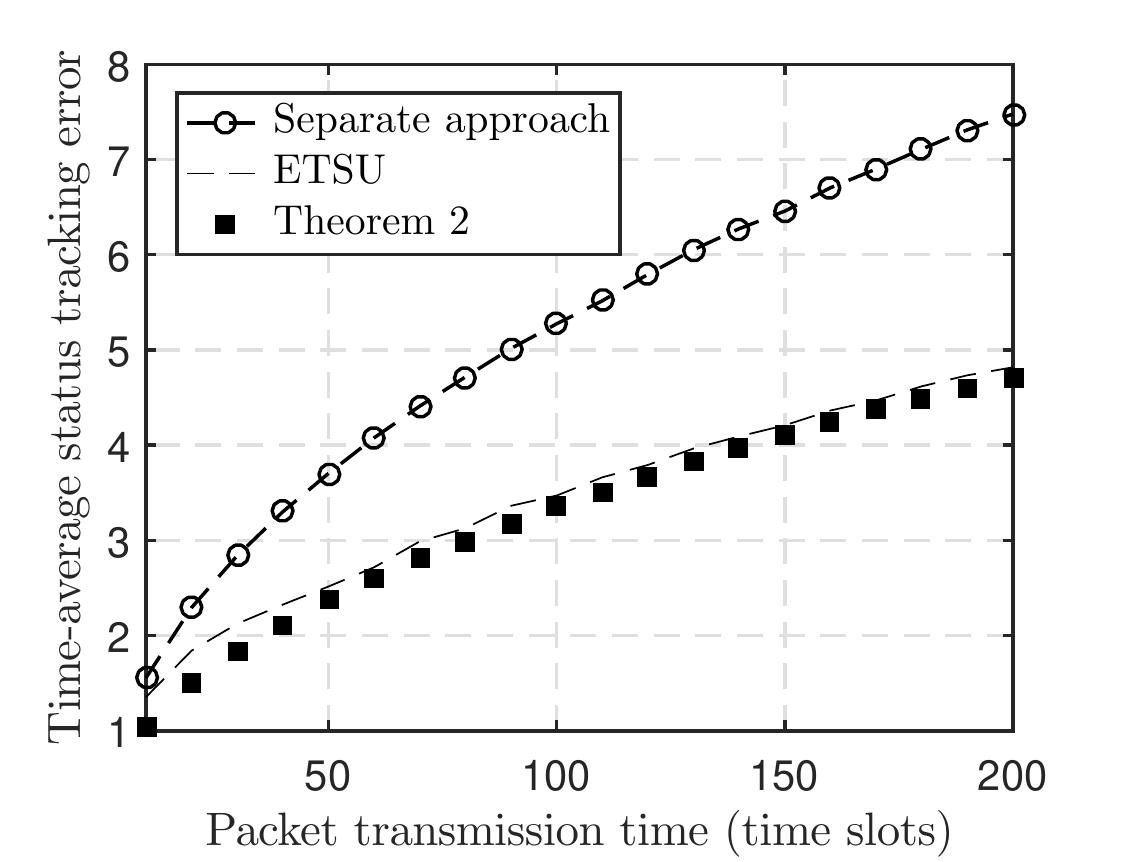}}
        \caption{Performance evaluations. (a) Theorem \ref{thm2} is compared with the optimum by MDP; (b) ETSU is compared with the centralized index policy (no random access) given by Theorem \ref{thm2}.}
        \label{fig_mdp}
    \end{figure}

    Compared with the centralized optimal status update scheme, the sub-optimality of ETSU comes from three possible aspects, which will be investigated in order in this, and following sections. First, the Whittle's index policy (Theorem \ref{thm2}), in contrast to the MDP optimal solution, is strictly speaking sub-optimal. Additionally, although the index in Theorem \ref{thm2} is precise, the index approximation in Corollary \ref{coro1} with unreliable channels needs to be validated. This aspect is evaluated in Fig. \ref{fig_mdp}(a), where it is observed that the approximate index with channel error, being derived assuming transmission failure probability approaches zero, is close to optimum with a wide range of $p_\mathsf{e}$. The $x$-axis is $p_{\mathsf{e},1}$ and we set $p_{\mathsf{e},2}=0.9$; we also let the error functions reflect the error sensitivities, i.e., $\delta_1(d) = d$ (error-tolerant) and $\delta_2(d) = e^d-1$ (error-sensitive).
    
    The second aspect of sub-optimality comes from the mean-field-based transmission probability mapping, which assumes the number of nodes is large. Fig. \ref{fig_mdp}(b) simulates scenarios with various numbers of nodes. Since it is impossible to obtain the MDP-based optimum with many nodes due to curse of dimensionality, the performance of ETSU is compared with centralized index policy by Theorem \ref{thm2} which has been shown near-optimal based on Fig. \ref{fig_mdp}(a). We adopt $\delta_n(d) = d$, $\forall n$, $\frac{t_{\mathsf{slot}}}{t_\mathsf{c}}=10$, and $p_{\mathsf{e},n}$ is uniformly generated from $[0,0.3]$. The transmission probability mapping function $\Psi(\cdot)$ is given by Corollary \ref{coro_mf} and we let $\nu=5/N$ such that there are about $5$ nodes with indices above the threshold at each time slot. In general, ETSU achieves near-optimal performance with a moderate number of nodes. Furthermore, the separate approach is also simulated with AoI-based scheduling (with channel error) according to index policy $I_{n}(h) = (1-p_{\mathsf{e},n})h_n^2$ derived in \cite{kadota18}; its performance is outperformed by ETSU in all cases.
    
    The third part is the modeling error of the information source nodes. We model them as random walk transitions, with several trivial extensions to, e.g., stay probability and asymmetrical walk discussed following \eqref{de}. Nevertheless, the current model cannot encompass state-inhomogeneous transitions which render the index policy untractable. In this regard, we test the algorithms based on real-world information sources in Section \ref{sec_sim}: CSI variations with an unmanned aerial vehicle (UAV) base station (BS) traversing an area. 
    \begin{figure}[!t]
        \centering
        \subfigure[Perfect]{\includegraphics[width=0.3\textwidth]{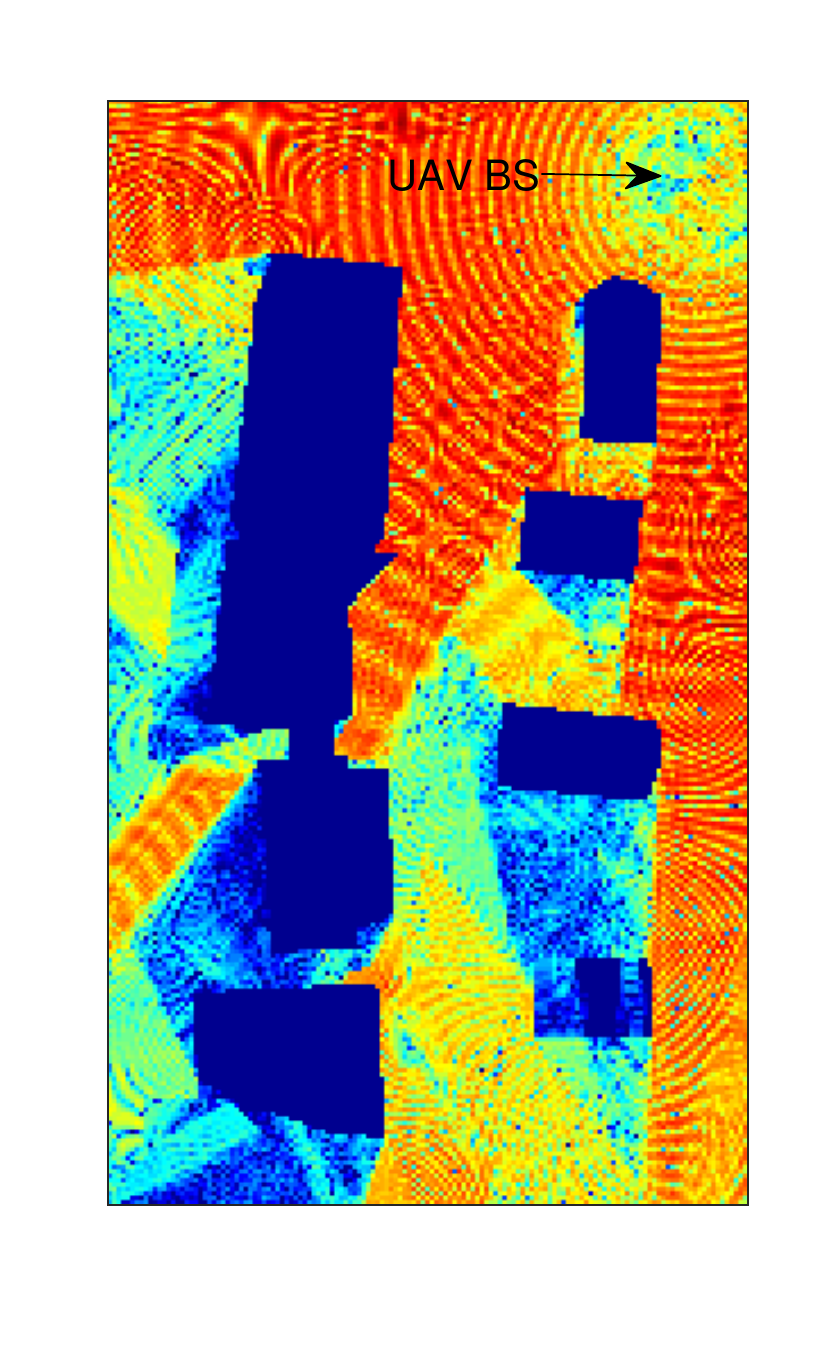}}
        \subfigure[Separate]{\includegraphics[width=0.3\textwidth]{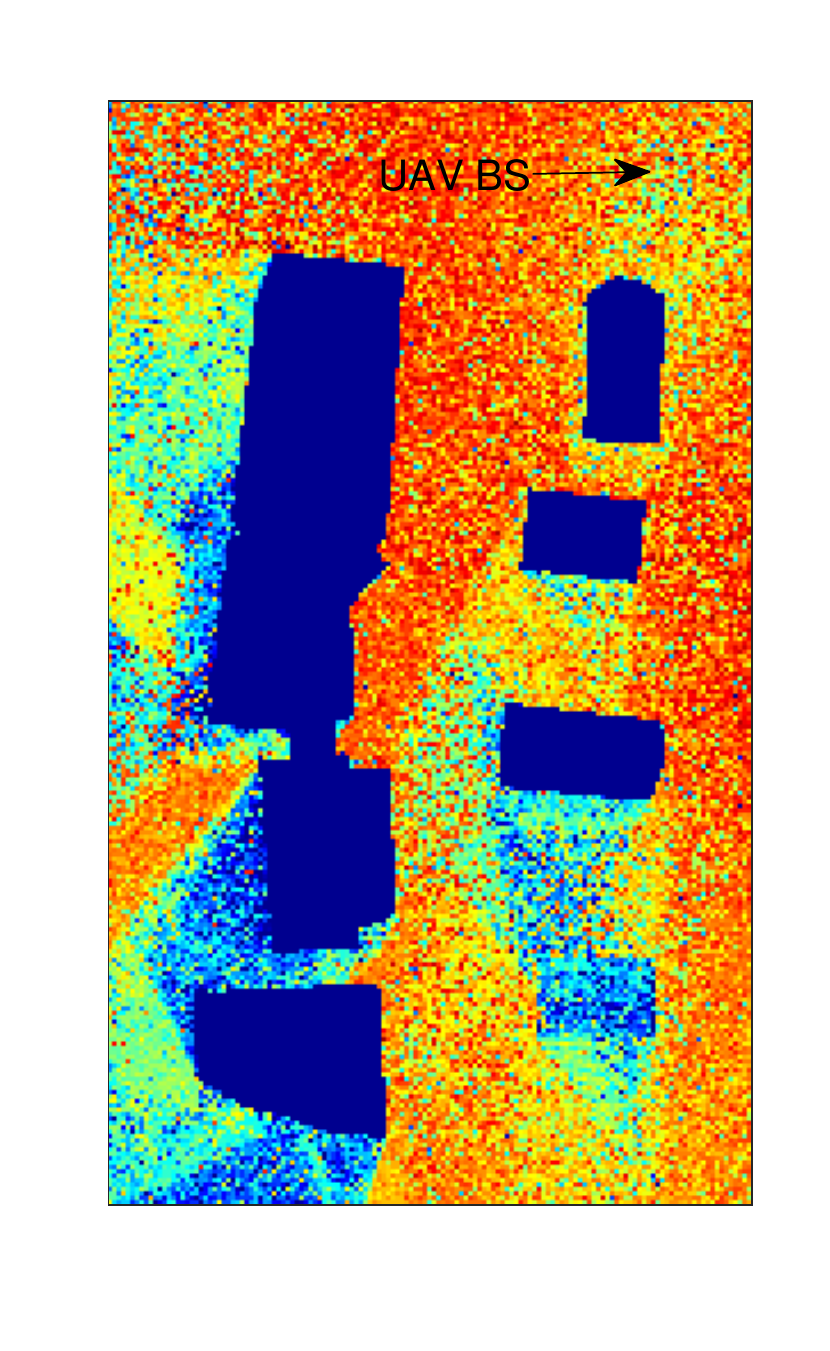}}
        \subfigure[ETSU]{\includegraphics[width=0.365\textwidth]{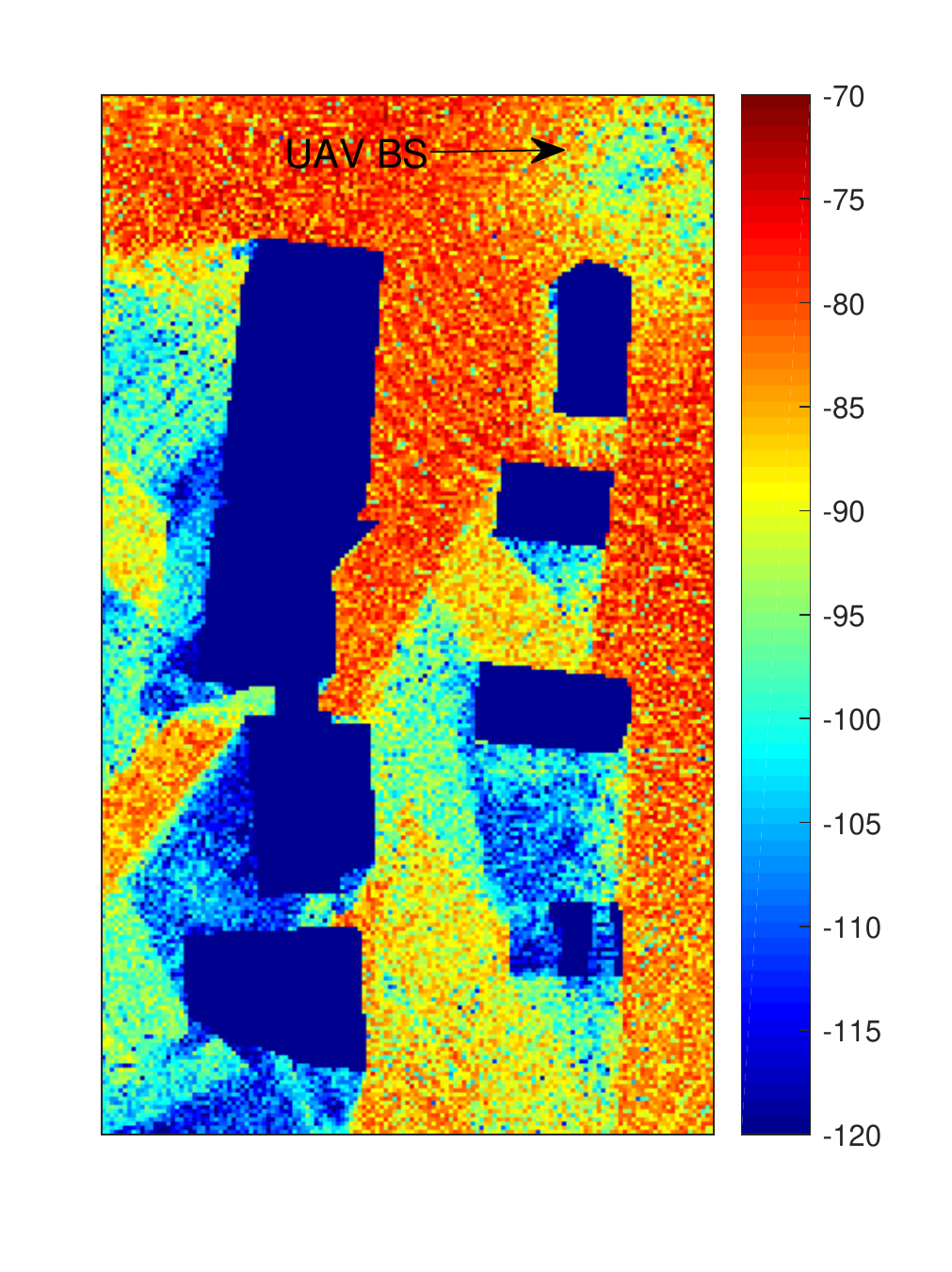}}
        \caption{CSI (in decibel-milliwatts) update based on ETSU and a separate approach \cite{jiang18_itc} with a UAV BS traversing over the area.}
        \label{fig_uav}
    \end{figure} 
    \section{Case Study: ETSU for Dynamic CSI Update}
    \label{sec_sim}
    We study a distributed CSI update use case where there are a large number of terminals ($1$~m apart in a $140$~m $\times$ $250$~m area and a total of $251\times141\approx3.5\times10^4$) reporting measured CSI to a sink node. The CSI variation is generated by the Wireless InSite$^{\circledR}$ ray-tracing simulator and letting a UAV BS (transmitter with a height of $50$~m) traverse the intended area from the bottom left corner to the top right with a velocity of $10$~km/h. We assume each CSI update occupies one time slot of length $0.5$~ms. The error function in ETSU is chosen as the squared error function, i.e., $\delta_n(s_{n}(t),\hat{s}_{n}(t)) = |s_{n}(t)-\hat{s}_{n}(t)|^2$. To be fair, we assume perfect CSI at the sink node at the beginning and evaluate the status tracking performance when the UAV flies past the area (at the top right corner). The CSI tracked at the sink node by a separate approach and ETSU is shown in Fig. \ref{fig_uav}, together with the perfect CSI. We assume reliable channel in this case. The separate approach based on \cite{jiang18_itc} implements sample-at-change sampling and scheduling that minimizes the average AoI of nodes, while neglecting the fact that the key status is not the one that is the stalest, but the one with, roughly speaking, the largest error. In contrast, ETSU, although assuming random-walk state transitions which does not match the CSI transitions, achieve evidently better CSI tracking performance.
    \section{Conclusions}
    \label{sec_con}
    A unified sampling and scheduling approach is proposed for status update in multiaccess wireless networks, capturing the key status variation in contrast to the conventional separate approach which samples the information sources and then schedules the nodes based on minimizing the AoI. The Whittle's index methodology is generalized to characterize the status packet importance based on arbitrary status tracking error functions. A mean-field approach is applied to derive the decentralized implementation in closed-form. As a special case of our results, we describe the closed-form decentralized non-linear AoI (arbitrary cost function of AoI) minimization scheme. The proposed ETSU evidently outperforms the separate approach as shown by extensive simulation results, including a realistic case study where we adopt ray-tracing generated CSIs as information sources with a mobile transmitter.
    
	\appendices
    \section{Proof of Theorem \ref{thm1}}
    \label{app_thm1}
    Define the cost-to-go function as $V_t({\mathcal{B}}(t))=\min_{\u(t),\u(t)^\mathsf{T}\boldsymbol{1}=1} q_t({\mathcal{B}}(t),\u(t))$, $q_t({\mathcal{B}}(t),\u(t)) \triangleq \mathbb{E}\left[V_{t+1}(\mathcal{B}(t+1))+g_t\left({\mathcal{B}}(t),\u(t)\right)\right]$, where $V_T(\mathcal{B}(T))=\mathbb{E}[g_T(\mathcal{B}(T))]$ is the terminal cost, the expected immediate cost at time $t$ is denoted by $\mathbb{E}[g_t\left(\mathcal{B}(t),\u(t)\right)]$, denote $\u(t) \triangleq [u_1(t),\cdots,u_N(t)]^\mathsf{T}\in\{0,1\}^N$, and the time index is omitted for brevity. According to the MDP methodology \cite{bersk}, the optimal total cost for the $T$-horizon problem is $V^{\ast}(\mathcal{B}(0))=V_0(\mathcal{B}(0))$, and the optimal dynamic policy is given by $\U^{\ast}=\{\u(0)^{\ast},\cdots,\u(T-1)^{\ast}\}$, where $\u(t)^{\ast}$ minimizes the cost-to-go for $t\in \{0,\cdots,T-1\}$. Because symmetrical two-state Markov sources are considered, it is clear that the system state can be simplified as $\hat{\mathcal{B}}(t) \triangleq \{\mathds{1}_{\{s_1(t)\neq\hat{s}_1(t)\}},\cdots,\mathds{1}_{\{s_N(t)\neq\hat{s}_N(t)\}}\}$, indicating whether the current statuses at source and destination are identical; we further denote $d_n(t) \triangleq \mathds{1}_{\{s_n(t)\neq\hat{s}_n(t)\}}$. 
	
	The proof includes two steps: first, we prove that the policy by \eqref{nc} is the \emph{myopic} policy, i.e., it minimizes the expected immediate cost (this is straightforward and hence omitted); secondly, a \emph{backwards induction} (on time $t$) based proof is given showing that the myopic policy is indeed the optimal policy. Define the total cost from time $t$ on given the current state and following the myopic policy \eqref{nc} as $W_t(\hat{\mathcal{B}}(t))$. Concretely, we will show the following statements are valid. Denote $\overline{W}^{t}_{n,m}(s,r) \triangleq W_t(c_1,\cdots,d_{n}(t)=s,\cdots,d_m(t)=r,\cdots,c_N)$ and $d_l(t)=c_l$, $\forall l\neq n,m$.
	
	\begin{itemize}
		\item[I.] $W_t(\hat{\mathcal{B}}(t))$ is the optimal total cost from $t$ on.
		\item[II.] For any $t \geq 0$ and $p_i\leq p_j$, $\overline{W}^{t}_{i,j}(0,1)\leq\overline{W}^{t}_{i,j}(1,0)$.
	\end{itemize}
	
	Let us first establish the induction basis. For $t=T$, hypothesis I is obviously true since \eqref{nc} is the myopic policy and the last step only concerns with the immediate cost. Hypothesis II also holds with equality.
	
	Suppose hypothesis I and II are valid from time $t+1$ to $T$, at time $t$ denote the myopic solution as $u_{n^\ast}(t)=1$ and zero otherwise, which indicates that $n^\ast = \argmin_{n\in \chi(t)} p_n$ where $\chi(t)$ is the set of nodes with $d_n(t)=1$. Then $\forall l\in \chi(t)$ with $p_l \ge p_{n^\ast}$, it follows from the cost-to-go function, after some manipulations, that
	\begin{iarray}
	&& q_t(\hat{\mathcal{B}}(t),n^\ast) - q_t(\hat{\mathcal{B}}(t),l) = \underbrace{\mathbb{E}[g_t\left(\mathcal{B}(t),n^\ast\right)] - \mathbb{E}[g_t\left(\mathcal{B}(t),l\right)]}_{\mathcal{M}_1} \nonumber\\
	&& + \sum_{c_n,n \neq l,n^\ast}\overline{p}_{n^\ast,l}\underbrace{\left(p_{n^\ast}-p_l)(\overline{W}^{t+1}_{n^\ast,l}(1,1)-\overline{W}^{t+1}_{n^\ast,l}(0,0))\right.}_{\mathcal{M}_2} \nonumber\\
	&& + \underbrace{\left.(1-p_{n^\ast}-p_l)(\overline{W}^{t+1}_{n^\ast,l}(0,1)-\overline{W}^{t+1}_{n^\ast,l}(1,0))\right)}_{\mathcal{M}_3},\nonumber
	\end{iarray}
    where $\overline{p}_{n^\ast,l}$ denotes the transition probability from $\hat{\mathcal{B}}(t)\backslash\{d_{n^\ast},d_l\}$ to $\{c_1,\cdots,c_N\}\backslash\{c_{n^\ast},c_l\}$, i.e., the transition probability to a given state excluding node $n^\ast$ and $c_l$ (without loss of generality, assume $n^\ast<l$). Note that it is unnecessary to consider nodes with index $m \notin \chi(t)$ since their states are correct. Based on the definition of myopic policy, $\mathcal{M}_1 \le 0$. It is straightforward that $\overline{W}^{t+1}_{n^\ast,l}(1,1)-\overline{W}^{t+1}_{n^\ast,l}(0,0) \ge 0$ since $\overline{W}^{t+1}_{n^\ast,l}(1,1)$ includes a state with two tracking errors and $\overline{W}^{t+1}_{n^\ast,l}(0,0)$ a state that corrects both; hence, combining with the $p_l \ge p_{n^\ast}$, we obtain $\mathcal{M}_2 \le 0$. Since $\forall n$, $p_n\le0.5$, and based on hypothesis II, we have $\mathcal{M}_3 \le 0$. Therefore, we arrive at the conclusion that the myopic solution at time $t$, i.e., $n^\ast$, is also the optimal action. With this, we have proved hypothesis I. Next, we can prove hypothesis II based on induction, by analyzing three cases on the myopic solution. The details are omitted due to lack of space. With this we conclude the induction proof.
    
    \section{Proof of Theorem \ref{thm2}}
    \label{app_thm2}
    Following the technique used in, e.g., \cite{kadota16,jiang18_itc,singh15}, we first assume the optimal policy has a threshold-type structure and solve the Bellman equations based upon it; after obtaining the optimal solution, consistency with this assumption is checked to conclude the proof. 
    
    Assume the optimal policy to solve \eqref{c2go} is update when $d \ge D$, and idle when $0 \le d < D$. Denote the upper term in the minimization in \eqref{c2go} as $\gamma_0(d)$ and the lower term $\gamma_1(d)$. Then for $d \ge D$, the optimal action is to update, and hence 
    \begin{iarray}
    \label{e0}
    f(d)=m+\frac{f(1)}{2}-\hat{J}^*, \, d \ge D.
    \end{iarray}
    Likewise, the optimal action is to idle otherwise, and 
    \begin{iarray}
    \label{diff}
    f(d) + \hat{J}^*=\delta(d)+\frac{1}{2}f(d+1)+\frac{1}{2}f(d-1), \, 0 < d < D, \nonumber
    \end{iarray}
    and the case with $d=0$ is specially treated with outcomes $f(1)=2\hat{J}^*$. Together with $f(0)=0$ we can obtain the formula for $f(d)$ when $0 \le d \le D$ based on the differential equation of \eqref{diff}.
    \begin{equation}
    \label{w1}
        f(d)=d(d+1) \hat{J}^* -2\sum_{i=1}^{d-1}\sum_{j=1}^{i}\delta(j),\, 0 \le d \le D.
    \end{equation}
    With the following three equations we can solve for the Whittle's index in \eqref{whittle}.
    \begin{iarray}
    \label{e1}
    f(D)&=&m,\\
    \label{e2}
    f(D)&=&D(D+1) \hat{J}^* -2\sum_{i=1}^{D-1}\sum_{j=1}^{i}\delta(j).\\
    \label{e3}
    \gamma_0(D) &=& \gamma_1(D).
    \end{iarray}
    Eq. \eqref{e1} and \eqref{e2} follow from \eqref{e0} and \eqref{w1} with $d=D$, respectively. Eq. \eqref{e3} is based on the fact that the auxiliary cost $m$ should be the minimum cost that make the update decision equally beneficial give the current state. Therefore, when $d=D$, two options should be equally valuable. 
    
    Consistency with the threshold-type assumption can be checked to be satisfied with the derived optimal policy, whose details are omitted for brevity. The indexability can be verified by checking that when $m=0$, the threshold is zero but update and idle are equally beneficial and hence one should idle; when $m \to \infty$, the threshold also goes to infinity. Additionally, the monotonicity can be proved by checking
    \begin{equation}
    I_{\mathsf{RW},n}(d)-I_{\mathsf{RW},n}(d-1)= d \delta_n(d)-\sum_{i=1}^{d-1} \delta_n(i) \overset{(a)}{\ge} \sum_{i=1}^d \delta_n(i)\ge0.\nonumber
    \end{equation}
    The inequality $(a)$ follows from the non-decreasing property of the error functions, i.e., $\delta_n(d) \ge \delta_n(i)$, $\forall i \le d$. With this, we conclude the proof.

    \section{Proof of Corollary \ref{coro1}}
    \label{app_coro1}
    The cost-to-go function is changed to
    \begin{iarray}
    \label{c2g2}
    f(d) + \hat{J}^*  &=&  \min \left\{ \begin{array}{l}
    \gamma_0(d),\\
    m + (1-p_{\mathsf{e}})f(1)/2 +p_{\mathsf{e}}\gamma_0(d)
    \end{array} \right\}, 
    \end{iarray}
    Following from \eqref{e3}, we can obtain $f(D)=\frac{m}{1-p_{\mathsf{e}}}$. For $d \ge D$, based on the lower term in \eqref{c2g2},
    \begin{equation}
    \frac{p_{\mathsf{e}}}{2} f(d+1) - f(d) + \frac{p_{\mathsf{e}}}{2} f(d-1) = p_{\mathsf{e}}\left(\hat{J}^*-\delta(d)\right)-m. \nonumber
    \end{equation}
    Solving this equation recursively gives us
    \begin{equation}
    \label{D1}
    p_{\mathsf{e}} f(D+1) - \left(1-\sqrt{1-p_{\mathsf{e}}^2}\right)f(D) = \frac{p_{\mathsf{e}}(m-p_{\mathsf{e}}\hat{J}^*)}{1-p_{\mathsf{e}}+\sqrt{1-p_{\mathsf{e}}^2}}.\nonumber
    \end{equation}
    Analyzing this in the regime $p_{\mathsf{e}} \to 0$, we can obtain 
    \begin{equation}
    \label{D2}
        f(D+1) \overset{p_{\mathsf{e}} \to 0}{\longrightarrow}\frac{m}{1-p_{\mathsf{e}}} + \smallO(1).
    \end{equation}
    Plugging \eqref{D2} into \eqref{c2g2}, we obtain a similar equation set as in \eqref{e1}-\eqref{e3}, expect that $m$ is replaced with $\frac{m}{1-p_{\mathsf{e}}}$, and hence the corresponding index follows immediately.     
    The indexability also follows because $\frac{m}{1-p_{\mathsf{e}}}$, compared with $m$, does not affect the monotonicity or index values at zero and infinity.
    \section{Proof of Theorem \ref{thm_mf}}
    \label{app_thm_mf}
    The state transition matrix with parameter $D_{\mathsf{th}}$ (denote $\gamma \triangleq 1-\lambda-\mu$) is $\boldsymbol{P}_{D_{\mathsf{th}}}\triangleq\left[
        \begin{array}{c|c}
        \overline{\boldsymbol{P}} & \tilde{\boldsymbol{P}} \\
        \hline
        \breve{\boldsymbol{P}} & \hat{\boldsymbol{P}}
        \end{array}
        \right] =$
    \begin{iarray}
    &&  \small        
        \left[
        \begin{array}{cccc|cccc}
        1-\lambda   & \lambda       &         &                    &  & & & \\
        \mu         & \gamma        & \lambda &                    &  & & & \\
                    & \ddots        & \ddots  &   \ddots           &  & & & \\
                    &               & \mu     & \gamma             & \lambda               &                         &&\\
        \hline
        \varepsilon &               &         & \mu(1-\varepsilon) & \gamma(1-\varepsilon) & \lambda(1-\varepsilon)  &&\\
        \vdots      &               &         &                    &         \ddots        &  \ddots                 & \ddots & \\        
        \end{array}
        \right], \nonumber
    \end{iarray}
    where $\overline{\boldsymbol{P}}$ is a $(D_{\mathsf{th}}-1)$-dimensional square matrix. Denote the stationary distribution as $\boldsymbol{\pi} \triangleq [\overline{\boldsymbol{\pi}}, \hat{\boldsymbol{\pi}}]$ correspondingly. Solving for the first $D_{\mathsf{th}}-1$ states yields (first assuming $\lambda \neq \mu$ and then generalized)
    \begin{equation}
    \label{11}
        \boldsymbol{\pi}_d = \left(\frac{\lambda}{\mu}\right)^d\left(\boldsymbol{\pi}_0-\frac{\varepsilon\sigma}{\lambda-\mu}\right)+\frac{\varepsilon\sigma}{\lambda-\mu},\,0 \le d < D_{\mathsf{th}},
    \end{equation}
    where $\sigma \triangleq \hat{\boldsymbol{\pi}} \boldsymbol{1}$. Solving for the remaining states yields
    \begin{equation}
    \label{12}
        \hat{\boldsymbol{\pi}} = \lambda \boldsymbol{\pi}_{D_{\mathsf{th}}-1} \boldsymbol{e}_1^{\mathsf{T}}\left(\boldsymbol{I}-\hat{\boldsymbol{P}}\right)^{-1},
    \end{equation}
    where $\boldsymbol{e}_1 \triangleq [1,0,0,\cdots]^{\mathsf{T}}$, and that $\boldsymbol{I}-\hat{\boldsymbol{P}}$ is invertable, since $\hat{\boldsymbol{P}}$ has eigenvalues smaller than one ($\varepsilon>0$). Plugging \eqref{11} into \eqref{12} yields
    \begin{equation}
    \label{13}
        \sigma = \lambda \beta \left[ \left(\frac{\lambda}{\mu}\right)^{D_{\mathsf{th}}-1}\left(\boldsymbol{\pi}_0-\frac{\varepsilon\sigma}{\lambda-\mu}\right)+\frac{\varepsilon\sigma}{\lambda-\mu}\right],
    \end{equation}
    where $\beta \triangleq \boldsymbol{e}_1^{\mathsf{T}}\left(\boldsymbol{I}-\hat{\boldsymbol{P}}\right)^{-1} \boldsymbol{1}$. Then summing over all states yields $\sum_{i=0}^{\infty} \boldsymbol{\pi}_i=1$, combining with \eqref{11} and \eqref{13} we obtain
    \begin{equation}
        \frac{1-\left(\frac{\lambda}{\mu}\right)^{D_{\mathsf{th}}}}{1-\frac{\lambda}{\mu}}\left(\boldsymbol{\pi}_0-\frac{\varepsilon \sigma}{\lambda-\mu} \right) + D_{\mathsf{th}} \frac{\varepsilon \sigma}{\lambda-\mu} + \sigma =1, \nonumber
    \end{equation}
    and hence $\boldsymbol{\pi}_0 = \left(\frac{\mu}{\lambda}\right)^{D_{\mathsf{th}}-1}\left(\frac{1}{\lambda \beta}-\frac{\varepsilon}{\lambda-\mu}\right)+\frac{\varepsilon \sigma}{\lambda-\mu}$. It follows that \eqref{48} gives the solution of $D_{\mathsf{th}}$. Next we will show that $\sigma$ is monotonically non-increasing with $D_{\mathsf{th}}$ and hence $\sigma=\nu$ yields a unique solution (note that when $D_{\mathsf{th}}=0$, $\sigma=1$; when $D_{\mathsf{th}} \to \infty$, $\sigma\to0$).
    \begin{iarray}
    \frac{\dx \sigma^{-1}}{\dx D_{\mathsf{th}}} &=& \frac{-\log\frac{\lambda}{\mu}}{\lambda\beta\left(1-\frac{\lambda}{\mu}\right)}\left(\frac{\lambda}{\mu}\right)^{1-D_{\mathsf{th}}} \nonumber\\
    && +\left(1+\frac{\log\frac{\lambda}{\mu}}{1-\frac{\lambda}{\mu}}\left(\frac{\lambda}{\mu}\right)^{1-D_\mathsf{th}}\right)\frac{\varepsilon }{\lambda-\mu} \nonumber\\
    &\overset{(a)}{\ge}& \left(1-\left(\frac{\lambda}{\mu}\right)^{1-D_\mathsf{th}}\right)\frac{\varepsilon }{\lambda-\mu} \ge 0,\nonumber
    \end{iarray}
    where the inequality $(a)$ stems from $\log(x) \le x-1$, $\forall x>0$. Thus far, we have shown that there is a unique solution to the threshold equation of \eqref{48}, and thereby the index threshold can be derived accordingly based on Theorem \ref{thm2}. Since the top $\nu$-fraction of nodes are allowed to compete for a transmission slot, which corresponds to approximately $ \nu N$ nodes, the transmission probability, i.e., $p_{\mathsf{tx}}$, can be derived based on the $p$-CSMA results in, e.g., \cite[Theorem 1]{gai11}.
    
	\bibliographystyle{ieeetr}
	\bibliography{re}
\end{document}